\documentclass[twocolumn,twoside,pra,superscriptaddress]{revtex4}
\usepackage[utf8]{inputenc}

\usepackage[pdftex]{graphicx,color}
\usepackage{amsmath}
\usepackage{amssymb}
\usepackage{epsfig}
\usepackage{amsthm}
\usepackage{mathtools}
\usepackage{tikz}
\usepackage{wrapfig}
\usepackage{comment}
\usepackage{simpler-wick}
\usepackage{simplewick}
\usepackage{latexsym,revsymb}
\usepackage[unicode=true,pdfusetitle,
bookmarks=true,bookmarksnumbered=false,bookmarksopen=false,
breaklinks=false,pdfborder={0 0 0},backref=false,colorlinks=false]
{hyperref}

\DeclareMathOperator*{\argmax}{argmax}

\newcommand{\ket}[1]{| #1 \rangle}

\newcommand{\vertiii}[1]{{\left\vert\kern-0.25ex\left\vert\kern-0.25ex\left\vert #1 
    \right\vert\kern-0.25ex\right\vert\kern-0.25ex\right\vert}}

\makeatletter
\newtheorem*{rep@theorem}{\rep@title}
\newcommand{\newreptheorem}[2]{%
\newenvironment{rep#1}[1]{%
 \def\rep@title{#2 \ref{##1}}%
 \begin{rep@theorem}}%
 {\end{rep@theorem}}}
\makeatother

\newreptheorem{theorem}{Theorem}

\newreptheorem{corollary}{Corollary}

\newreptheorem{lemma}{Lemma}

\begin{document}
\title{Local tensor-network codes}
\author{Terry Farrelly}
\email{farreltc@tcd.ie}
\affiliation{ARC Centre for Engineered Quantum Systems, School of Mathematics and Physics, The University of Queensland, St Lucia, QLD, 4072, Australia}
\author{David K. Tuckett}
\affiliation{Centre for Engineered Quantum Systems, School of Physics,
University of Sydney, Sydney, NSW 2006, Australia}
\author{Thomas M. Stace}
\affiliation{ARC Centre for Engineered Quantum Systems, School of Mathematics and Physics, The University of Queensland, St Lucia, QLD, 4072, Australia}
\begin{abstract}
	Tensor-network codes enable the construction of large stabilizer codes out of tensors describing smaller stabilizer codes.  An application of tensor-network codes was an efficient and exact decoder for holographic codes.
	Here, we show how to write some topological codes, including the surface code and colour code, as simple tensor-network codes.
	We also show how to calculate distances of stabilizer codes by contracting a tensor network.  The algorithm actually gives more information, including a histogram of all logical coset weights.  We prove that this method is efficient in the case of holographic codes.  Using our tensor-network distance calculator, we find a modification of the rotated surface code that has the same distance but fewer minimum-weight logical operators by injecting the non-CSS five-qubit code tensor into the tensor network.  This corresponds to an improvement in successful error correction of up to $2\%$ against depolarizing noise (in the perfect-measurement setting), but comes at the cost of introducing four higher-weight stabilizers.
	Our general construction lets us pick a network geometry (e.g., a Euclidean lattice in the case of the surface code), and, using only a small set of seed codes (constituent tensors), build extensive codes with the potential for optimisation.
\end{abstract}

\maketitle

\section{Introduction}
Tensor networks \cite{BC17} have proved useful in quantum error correction, primarily for decoding \cite{FP14,BSV14,FP14a,Darmawan17,Darmawan18,CF18,FHM20,Chubb21}.  In \cite{FP14}, Ferris and Poulin created a decoder by representing the encoding circuit of the code as a tensor network for circuits that were similar to the MERA tensor network.  Using a different approach, the surface code was decoded by contracting a two-dimensional tensor network in \cite{BSV14}.  This method was applied to several different noise models \cite{CF18,TDC19,TBF18} and also expanded to general codes with local stabilizers in two dimensions \cite{Chubb21}.  Non unitary and correlated noise were decoded for the surface code using a PEPO representation of the state \cite{Darmawan17,Darmawan18}.  Tensor networks have also been instrumental for constructing holographic error correcting codes \cite{FYH15,LS15,HNQ16,Evenbly17,HMBS18,BO18,JGP19,JGP19a,KC19,OS20,MFG20}.

More recently, ``tensor-network codes'' \cite{FHM20,FHM20b} were introduced, which use tensors to describe stabilizer codes.  This allows one to easily construct new larger codes by connecting many tensors (i.e., by contracting tensor indices) to form a new code.  Trivial examples are concatenated and convolutional codes, but more interesting examples include holographic codes \cite{FHM20}.  In principle, there is no restriction on the geometry of the tensor network, though some geometries lend themselves to efficient tensor contraction, which naturally produces an efficient decoder \cite{FHM20}.

In this work, we build up more new tensor-network codes.  For example, we look at topological codes, where we see that surface codes are very simple examples of tensor-network codes, as is the colour code.  We also show how to use tensor networks to calculate distances for stabilizer codes.  After first applying this to some simple examples of stabilizer codes, we then use this distance calculator to modify the surface code to find a code with the same distance but with fewer minimal-weight logical operators.  This modified code has an improved success probability of correcting depolarizing noise of up to $2\%$.  This works simply by substituting the (non CSS) five-qubit code tensor into the tensor network describing the surface code.

\section{Stabilizer codes}
In this paper, we focus on stabilizer codes \cite{Gottesman97,Gottesman09,NielsenChuang,Roffe19}, in which stabilizers and logical operators are elements of $\mathcal{G}_n$, which is the $n$-qubit Pauli group.  The group $\mathcal{G}_n$ comprises all operators of the form $z\sigma^{i_1}\otimes...\otimes\sigma^{i_n}$, with $z\in\{\pm 1,\pm i\}$.  Here, we are denoting the identity and three Pauli operators via $\sigma^{0}=\openone$, $\sigma^{1}=X$, $\sigma^{2}=Y$ and $\sigma^{3}=Z$.

The codespace, which is the subspace of the Hilbert space used to encode logical information, is fully determined by the stabilizers, an abelian group of Pauli operators, i.e., $\mathcal{S}\subset\mathcal{G}_n$.  The codespace is determined by the fact that every state $\ket{\psi}$ in the codespace satisfies $S\ket{\psi}=\ket{\psi}$ for all stabilizers $S\in\mathcal{S}$.  If we denote the number of independent generators of $\mathcal{S}$ by $r$, then the dimension of the codespace is $2^{n-r}$, which corresponds to $k=n-r$ encoded logical qubits \cite{NielsenChuang}.  We denote generators by $S_i$.

The logical operators of the code generate a group $\mathcal{L}\subset\mathcal{G}_n$, which is non-abelian.  One set of generators for the group is the $k$ $Z$-type and $k$ $X$-type operators.  (By $Z$-type and $X$-type, we mean that these operators are canonically conjugate to each other, not that they only consist of $\sigma^1$ or $\sigma^3$ Pauli operators.)  We denote these by $Z_{\alpha}$ and $X_{\alpha}$ respectively, with $\alpha\in\{1,..,k\}$.  These operators commute with all stabilizers, while anticommuting pairwise, i.e., $X_{\alpha}Z_{\beta}=(-1)^{\delta_{\alpha \beta}}Z_{\beta}X_{\alpha}$.

It will be useful to also consider the (abelian) group of operators called pure errors (a.k.a.\ destabilizers) $\mathcal{P}\subset\mathcal{G}_n$.  This group is defined by $n-k$ operators $P_i$ which satisfy $P_iS_j=(-1)^{\delta_{ij}}S_jP_i$ and $P^2_i=\openone$.  The Pauli group on $n$ physical qubits is generated by products of $P_i$, $S_i$, $X_{\alpha}$ and $Z_{\alpha}$.

\begin{table}
\begin{center}
  \begin{tabular}{| c | c c c c c c | }
    \hline
    Qubit & $0$ & 1 & 2 & 3 & 4 & 5 \\ \hline
    $S_1$ & $\openone$ & $X$ & $Z$ & $Z$ & $X$ & $\openone$ \\ \hline
    $S_2$ & $\openone$ & $\openone$ & $X$ & $Z$ & $Z$ & $X$ \\ \hline
    $S_3$ & $\openone$ & $X$ & $\openone$ & $X$ & $Z$ & $Z$ \\ \hline
    $S_4$ & $\openone$ & $Z$ & $X$ & $\openone$ & $X$ & $Z$ \\ \hline
    $X_1$ & $X$ & $X$ & $X$ & $X$ & $X$ & $X$ \\ \hline
    $Z_1$ & $Z$ & $Z$ & $Z$ & $Z$ & $Z$ & $Z$ \\ 
    \hline
  \end{tabular}
\end{center}
	\caption{Logical operators and stabilizer generators for the five-qubit code.  The operators on physical qubits are shown in columns $1-5$, and the action on the codespace is in column $0$.  We can think of all six operators on $0-6$ as stabilizing a six-qubit stabilizer state, i.e., a purified five-qubit code.}
\label{table:five-qubit}
\end{table}

Finally, during error correction, we measure each of the stabilizer generators (which have eigenvalues $\pm 1$), giving the error syndrome.  This is a length-$r$ vector $\vec{s}$ of the measurement outcomes, with $s_i=0$ if the error commuted and $s_i=1$ if the error anticommuted with stabilizer $S_i$.  Finding the optimal correction operator given the syndrome is extremely difficult in general \cite{HG11,IP13}, though some algorithms work well for some codes (e.g., minimum-weight perfect matching for the surface code).  It is always straightforward to find \textit{an} error consistent with a given syndrome by using the pure errors, i.e., the operator $P(\vec{s}\,)=\prod_i P^{s_i}_i$ gives rise to the syndrome $\vec{s}$.  But quantum codes are degenerate, meaning multiple different errors can have the same syndrome, e.g., $P(\vec{s}\,)$ and $P(\vec{s}\,)S$ for any stabilizer $S$ will give rise to the same error syndrome.

\section{Stabilizer codes from tensor networks}
As introduced in \cite{FHM20}, we represent small ``seed'' stabilizer codes by tensors and use them to build up bigger stabilizer codes.  This works as follows.  We can represent Pauli operators by strings of integers, so the operator $XZZX\openone=\sigma^{1}\sigma^{3}\sigma^{3}\sigma^{1}\sigma^{0}$ would be represented by the string $(1,3,3,1,0)$.

For a stabilizer code with $n$ physical and $k$ logical qubits, we define the rank-$n$ tensor 
\begin{equation}\label{eq:Tdef}
	T(L)_{g_1,...,g_n}=\begin{cases}
                     1 \mathrm{\ \ \  if\ }\sigma^{g_1}\otimes...\otimes\sigma^{g_n}\in \mathcal{S}L\\
                     0 \mathrm{\ \ \  otherwise,}
                    \end{cases}
\end{equation}
for each logical operator $L$.  Here the indices $g_j\in\{0,1,2,3\}$, and $\mathcal{S}L$ is the set of all operators $SL$ with $S\in\mathcal{S}$.  In other words, $\mathcal{S}L$ is the coset of $\mathcal{S}$ with respect to logical $L$.  We can also think of $T(L)_{g_1,...,g_n}$ as the indicator function for all operators in the class $L$.

As an example, $T(\openone)_{g_1,...,g_n}$ describes the stabilizer group, as it is non zero only if $\sigma^{g_1}\otimes...\otimes\sigma^{g_n}$ corresponds to a stabilizer.  This does not include the signs of the stabilizers, though once these are fixed for a set of generators, then they are fully determined for the rest of the group.  Also, for a stabilizer code that has only a single logical qubit, the tensor $T(X)_{g_1,...,g_n}$ describes the coset corresponding to the logical $X$ operator.

As explained in \cite{FHM20}, contracting seed tensors generates larger codes, which come with a natural tensor-network decoder.  A useful idea then is to start with a tensor-network geometry that is efficiently contractible and use this to construct codes that we know we can decode efficiently since we can contract the tensor network.  We can then vary the code tensors within this geometry to search for good codes.

\begin{figure*}[ht!]
\includegraphics[width=\textwidth]{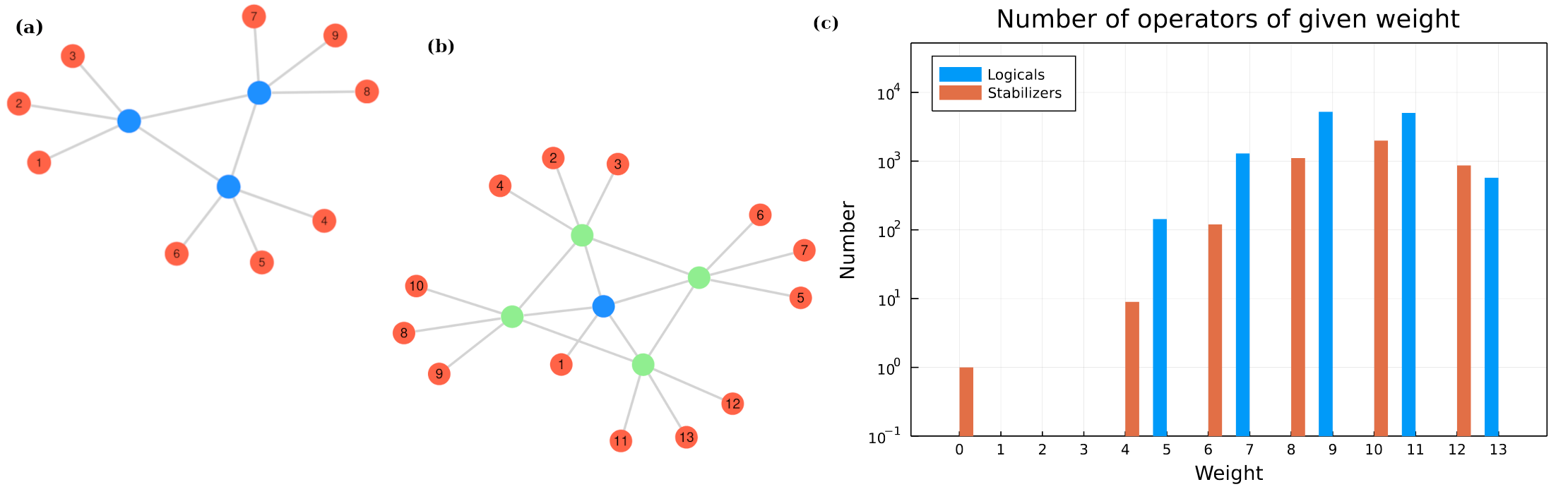}
	\caption{Example tensor-network code graphs.  Code tensors introducing one logical qubit per code are shown in blue, code tensors with no logicals are green, and physical qubits are shown in red.
	(a) Each code tensor is a five-qubit code tensor, which gives rise to a $[[8,3,3]]$ code.
	(b) shows a larger tensor-network code with $13$ physical qubits and one logical qubit.  The central blue tensor is a five-qubit code tensor, and the remaining code tensors are purified five-qubit code tensors.
	(c) Histogram of operator weights for the code in (b).  Stabilizers (the identity coset) are shown in red and logical operators (not including the identity) are shown in blue.  Note that only one bar (red or blue) appears for, e.g., weight-five operators because there are no weight-five stabilizers, only weight-five logical operators.  Here we see that the code in (b) is a $[[13,1,5]]$ code.  (In fact, contracting indices corresponding to qubits $2$ and $3$, generates a $[[11,1,5]]$ code, which is the maximum distance possible with $11$ physical qubits and one logical qubit \cite{Grassl21}.)
	}
\label{fig:example}
\end{figure*}

To see how to contract code tensors to get new stabilizer codes is straightforward.  First, the tensor describing a tensor product of two stabilizer codes is just the product of the code tensors:
\begin{equation}
 \begin{split}
	 & T_{12}(L_1\otimes L_2)_{g_1,...,g_{n_1},h_1,...,h_{n_2}} = \\
	 & T_1(L_1)_{g_1,...,g_{n_1}}T_2(L_2)_{h_1,...,h_{n_2}}.
\end{split}
 \end{equation}
 Second, we contract two indices (say $g_1$ and $h_1$ above).  To understand when the resulting tensor describes a bona fide stabilizer code, we look at the more general case of contracting two indices of a code tensor.  Take code tensor $T(L)_{j_1,...,j_{n}}$ with $n$ physical qubits and $k$ logical qubits.  Suppose we contract two indices (take the first two indices for simplicity), i.e.,
 \begin{equation}
	 T_{\mathrm{new}}(L)_{j_3,...,j_{n}} = \sum_{i\in\{0,1,2,3\}}\!\!T(L)_{i,i,j_3,...,j_{n}}.
 \end{equation}
 It is straightforward to check if the left hand side is a code tensor describing a stabilizer code, as shown in more detail in appendix \ref{sec:contraction}.  In short, a sufficient condition ensuring that the resulting tensor also describes a stabilizer code is if we can find two stabilizers of the original code $S_1$ and $S_2$ that have the following form.  Writing $S_1=\sigma^{j_1}\otimes\sigma^{j_2}\otimes\sigma^{j_3}\otimes...$ and $S_2=\sigma^{i_1}\otimes\sigma^{i_2}\otimes\sigma^{i_3}\otimes...$, we require that $i_1 \neq i_2$ and $j_1 \neq j_2$.  We also require that after taking their product $S_1S_2=\sigma^{k_1}\otimes\sigma^{k_2}\otimes\sigma^{k_3}\otimes...$ with $k_1\neq k_2$.  This condition is sufficient but not necessary in some cases, as explained in appendix \ref{sec:contraction}.  From a numerical perspective, two stabilizers satisfying the conditions above can be found quickly (in $O(n)$ time), and stabilizer generators and logical operators for the new stabilizer code can also be found quickly (also in $O(n)$ time).

With this formalism it is straightforward to build larger error correcting codes using smaller seed code tensors as building blocks.  These tensor-network codes include many holographic codes as well as
concatenated codes and generalized concatenated codes \cite{GSS09,WZG13}.  However, there is no restriction on the geometry of the network or the number of logical qubits.

Physically, in a photonic setting we can think of the contraction of two tensor indices as type-II fusion \cite{BR05,BBB21}, where we are measuring $\sigma^1\otimes\sigma^1\otimes\openone...$ and $\sigma^3\otimes\sigma^3\otimes\openone...$.  

In Figure \ref{fig:example} (a), we give an example of a simple tensor-network code composed of three five-qubit code tensors (see Table \ref{table:five-qubit} for a description of the five-qubit code).  This is a $[[8,3,3]]$ code, which is the maximum distance for a code with eight physical and three logical qubits \cite{Grassl21}.

\section{Calculating code distance via tensor networks}
The tensor-network code construction naturally gives rise to an algorithm for computing the code distance.  In fact, we can calculate more information:\ by contracting a tensor network, we generate a histogram of operators (logicals and/or stabilizers) according to their weight.  If the tensor-network is efficiently contractible, then this histogram can be generated efficiently.

To see this, start by defining a weight tensor:
\begin{equation}
\label{eq:weighttens}
 W_w^{g_1,...,g_n}=\begin{cases}
                    1\ \mathrm{if}\ \textsf{weight}(\sigma^{g_1}\otimes...\otimes\sigma^{g_n}) = w\\
                    0\ \mathrm{otherwise},
                   \end{cases}
\end{equation}
where the function $\textsf{weight}$ returns the weight of a Pauli operator, i.e., the number of locations where it acts non trivially.  To find the distribution of operator weights of a tensor-network code, we use the representation $T^{l_1,...,l_k}_{g_1,...,g_n}=T(L)_{g_1,...,g_n}$, where $l_i=0,1,2,3$ indicates the logical identity, $X$, $Y$ and $Z$ operator on the $i$th logical qubit.  Then the number of operators of each weight for each logical coset is given by
\begin{equation}
 C_w^{l_1,...,l_k} = W_{w}^{g_1,...,g_n}T^{l_1,...,l_k}_{g_1,...,g_n},
\end{equation}
where repeated indices are summed.  For example, $C_w^{0,...,0}$ is a list of the numbers of stabilizers of each weight.  Then we calculate the distance of the code using
\begin{equation}
 D_w = \sum_{l_1,...,l_k}(C_w^{l_1,...,l_k})-C_w^{0,...,0}.
\end{equation}
Here we sum over all logical cosets and then subtract the identity coset (the stabilizer group).  The code distance is then given by the first value of the index $w$ for which $D_w \neq 0$.

On its own, this is not particularly useful, as the tensors $T$ or $W$ will become unwieldy for large $n$.  However, we split $W$ into a product of smaller tensors by first generalizing it:
\begin{equation}
 W_{w_{\mathrm{out}},w_{\mathrm{in}}}^{g_1,...,g_n}=\begin{cases}
                    1\ \mathrm{if}\ \textsf{weight}(\sigma^{g_1}\otimes...\otimes\sigma^{g_n}) = w_{\mathrm{out}}-w_{\mathrm{in}}\\
                    0\ \mathrm{otherwise}.
                   \end{cases}
\end{equation}
We note that $W_{w_{\mathrm{out}}}^{g_1,...,g_n} = W_{w_{\mathrm{out}},0}^{g_1,...,g_n}$.  It follows that these tensors can be chained in a natural way.
\begin{equation}
 W_{w_{3},w_{1}}^{g_1,...,g_n,h_1,...,h_m}=\sum_{w_2}W_{w_{3},w_{2}}^{g_1,...,g_n}W_{w_{2},w_{1}}^{h_1,...,h_m}.
\end{equation}
We can chain these tensors in a way that mimics the tensor-network geometry of the original code.  As a result if the original code tensor network can be contracted easily, so can the tensor network to calculate weights.

\begin{figure*}[ht!]
\includegraphics[width=\columnwidth]{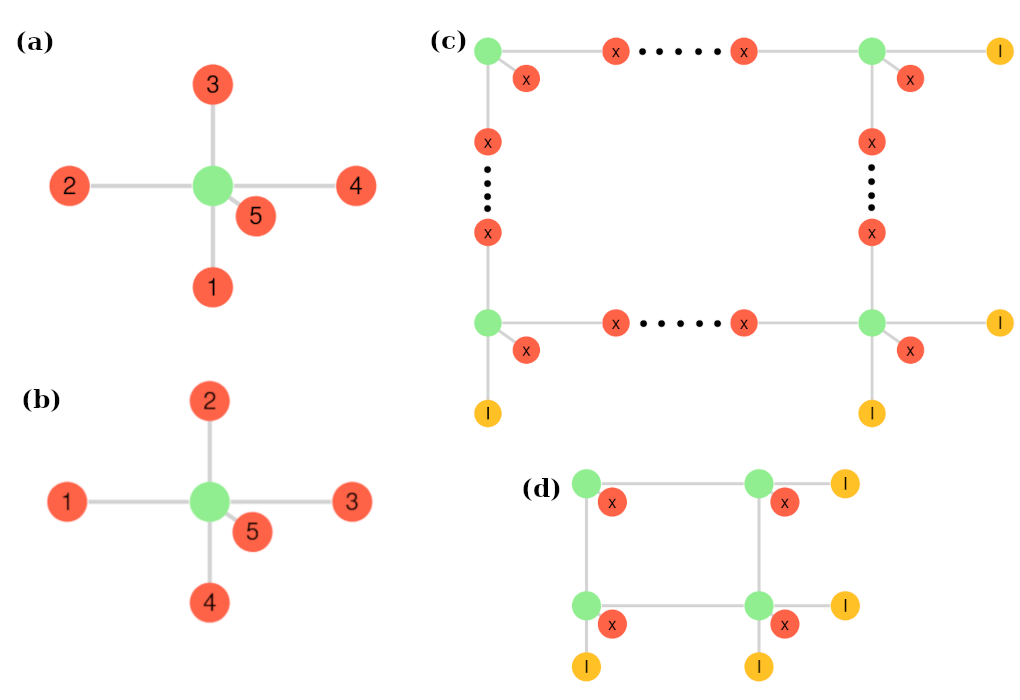}
\includegraphics[width=\columnwidth]{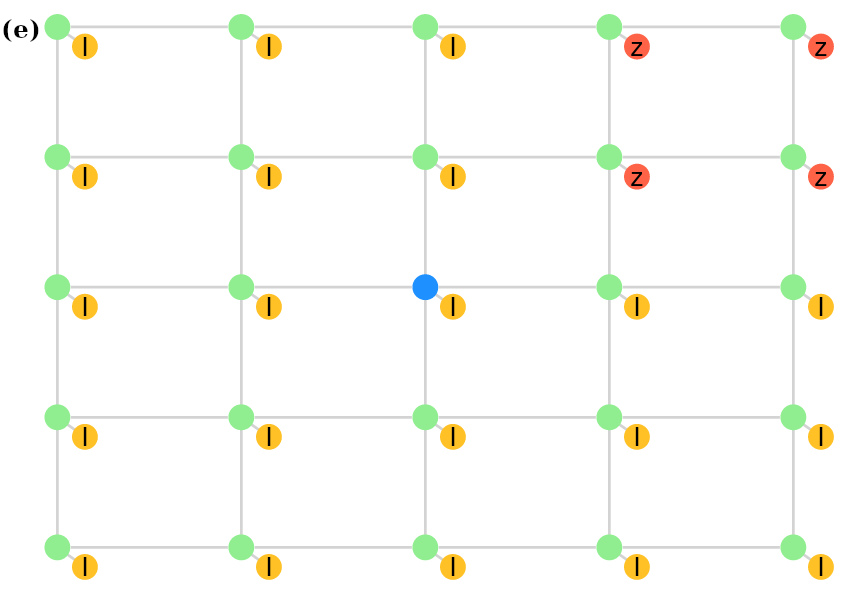}
	\caption{(a) \& (b) show bulk tensors for the surface code.  Numbers correspond to qubits in Table \ref{table:surf5}.  The code tensor in (a) includes logical $X$ as an additional stabilizer, while the code tensor in (b) includes logical $Z$ as a code stabilizer. (c) shows stabilizers of code tensors [at the top left corner of the tensor network in (e)] before contraction of indices, and (d) shows the resulting stabilizer after contraction of tensor indices.  (e) shows the $[[25,1,5]]$ rotated planar surface code as a tensor-network code.  Each green tensor in the bulk is a five-qubit surface code tensor with alternating tensors having logical $X$ or $Z$ as an additional stabilizer as shown in (a) \& (b).  The central tensor (blue) has a single logical qubit.  Yellow and red circles represent physical qubits.  Here we see a plaquette stabilizer, with non-trivial Paulis in red, and identities in yellow.  Our modified code simply replaces the blue tensor, which corresponded to the five-qubit \textit{surface} code, by the non-CSS five-qubit code (the ordering of the legs was chosen to match the surface code's distance, while lowering the number of minimal-weight logical operators).}
\label{fig:surface}
\end{figure*}

Let us consider a simple example.  As shown in Figure \ref{fig:example} (b), we have contracted five code tensors, with one corresponding to the five-qubit code, and the rest corresponding to the purified five-qubit code (see Table \ref{table:five-qubit}), so there is a single logical qubit.  By using our tensor-network method, we find the distribution of operator weights as shown in figure \ref{fig:example} (c).  We see that the distance is five, so this is a $[[13,1,5]]$ code.

This is a promising new way to find code distances, provided we can contract the tensor network.  Then a useful idea is to start with tensor-network geometries that we know we can easily contract and then iterate over different choices of seed code tensors to find high-distance codes.  We can also look for codes that not only have high distance but also have many low-weight stabilizers.  This method can also be expanded to find the distribution of logical operators by weight restricted to those of a certain form, e.g., those that have only identity and Pauli $Z$ components.  This can be used to identify codes that are tailored to biased noise \cite{TDC19}.

Let us end this section by noting that this distance calculation method is efficient for holographic codes (see appendix \ref{sec:holo}), as well as for any code with simpler tensor networks, such as trees.

\section{Maximum likelihood decoding via tensor networks}
Maximum likelihood decoding is optimal, as it uses the syndrome to find the correction operator that is most likely to return the code to the correct code state.  For general quantum codes, this is a very difficult computational task \cite{IP13}.  One method that works well for some codes uses tensor networks \cite{FP14,BSV14,CF18,FHM20,Chubb21}.

We can write any error with syndrome $\vec{s}$ as $P(\vec{s}\,)SL$, where $P(\vec{s}\,)$ is the pure error corresponding to syndrome $\vec{s}$, and some $S\in\mathcal{S}$ and $L\in\mathcal{L}$.  The key point is that $P(\vec{s}\,)SL$ has the same effect on the codespace regardless of which stabilizer $S$ appears.  In contrast, different logicals $L$ will have different effects on the codespace.  To find the optimal correction, we need to calculate
\begin{equation}
 \chi(L,\vec{s}\,)=\sum_{S\in\mathcal{S}} \mathrm{prob}(P(\vec{s}\,)SL)
\end{equation}
for each logical $L\in\mathcal{L}$.  Here $\mathrm{prob}(P(\vec{s}\,)SL)$ is the probability that the error $P(\vec{s}\,)SL$ occurred.  Then the best correction operator is $\overline{L}P(\vec{s}\,)$, where $\overline{L}=\argmax_L\chi(L,\vec{s}\,)$.

Using tensor-network codes,
we should think of the noise distribution $\mathrm{prob}(P(\vec{s}\,)SL)$ as a tensor.  To do this, we write $SL=\sigma^{g_1}\otimes...\otimes\sigma^{g_n}$ and $P(\vec{s}\,)=\sigma^{e_1}\otimes...\otimes\sigma^{e_n}$.  Recall that $P(\vec{s}\,)$ is fixed by the syndrome.  Then we define
\begin{equation}
 \mathcal{E}(\vec{s}\,)^{g_1,...,g_n}= \mathrm{prob}(\sigma^{e_1}\sigma^{g_1}\otimes...\otimes\sigma^{e_n}\sigma^{g_n}).
\end{equation}
For uncorrelated error models (e.g., i.i.d.\ depolarizing noise), this factorizes simply to a product of rank-one tensors:\ $\mathcal{E}(\vec{s}\,)^{g_1,...,g_n}  = p_1(\sigma^{e_1}\sigma^{g_1})\times...\times p_n(\sigma^{e_n}\sigma^{g_n})$, with $p_i(\sigma^{a_i})$ being the probability that the noise will cause $\sigma^{a_i}$ to act on qubit $i$.

Calculating $\chi(L,\vec{s}\,)$ involves contracting the code tensor with $\mathcal{E}$ to get
\begin{equation}
 \chi(L,\vec{s}\,)=T(L)_{g_1,...,g_n}\mathcal{E}(\vec{s}\,)^{g_1,...,g_n}.
\end{equation}
As we saw for the tensor-network distance calculator, this is useful if $T$ and $\mathcal{E}(\vec{s}\,)$ both factorize in a way that is amenable to contraction.  This is made especially easy for i.i.d.\ noise, but correlated noise can also be handled in some cases if, e.g., it has a finite correlation length, as is the case for factored noise \cite{CF18}.

There are some important examples where this decoding approach works well, such as (planar) surface codes \cite{BSV14}, codes based on the MERA tensor network \cite{FP14}, and holographic codes \cite{FHM20}.  It is also clear that such an approach would work well for concatenated or convolutional codes, but these already have efficient and exact decoders.  Typically, this decoding method involves approximations, e.g., in \cite{BSV14} the two-dimensional tensor network was contracted using the MPS-MPO method, which uses bond-dimension truncation.

\section{Topological codes as tensor-network codes and modifications}
\label{sec:top}
We view several important topological codes as tensor-network codes built from smaller codes. For example, the rotated surface code can be seen as a tensor-network code made of surface code fragments.  This is constructed from bulk tensors corresponding to the five-qubit surface code with stabilizers in Table \ref{table:surf5}.  The construction is most easily understood from Figure \ref{fig:surface}.  Similarly one can construct the colour code, as described in appendix \ref{sec:colour}, and the original (unrotated) surface code.

\begin{table}
\begin{minipage}{0.5\linewidth}
		\centering
    \begin{tabular}{| c | c c c c c | }
    \hline
    Qubit & 1 & 2 & 3 & 4 & 5 \\ \hline
    $S_1$ & $Z$ & $Z$ & $\openone$ & $\openone$ & $Z$  \\ \hline
    $S_2$ & $\openone$ & $\openone$ & $Z$ & $Z$ & $Z$  \\ \hline
    $S_3$ & $\openone$ & $X$  & $X$ & $\openone$ & $X$  \\ \hline
    $S_4$ & $X$ & $\openone$ & $\openone$ & $X$ & $X$ \\ \hline
    $Z_1$ & $Z$ & $\openone$ & $\openone$ & $Z$ & $\openone$  \\ \hline
    $X_1$ & $X$ & $X$ & $\openone$ & $\openone$ & $\openone$  \\ 
    \hline
  \end{tabular}
	\end{minipage}
	\begin{minipage}{0.45\linewidth}
		\centering
		\includegraphics[scale=0.35]{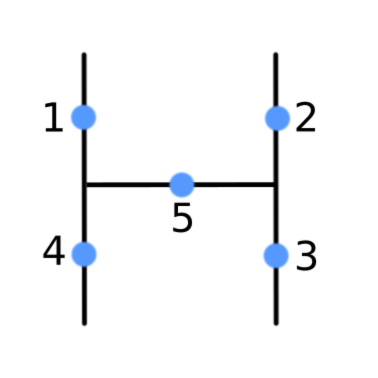}
	\end{minipage}
	\caption{Logical operators and stabilizer generators for the $[[5,1,2]]$ surface code fragment.}
\label{table:surf5}
\end{table}

Consider a rotated $[[L^2,1,L]]$ surface code with length $L$.  
The degeneracy of the lowest-weight logical operators is actually quite high and grows quickly with $L$, e.g., for $L=7$ it is already $896$, versus only $14$ for the original surface code with $L=7$.  It seems natural that keeping distance and code size fixed, while lowering the degeneracy of the lowest-weight logical operators may be an interesting avenue to explore to find better performing codes.

\begin{figure}[ht!]
\includegraphics[width=\columnwidth]{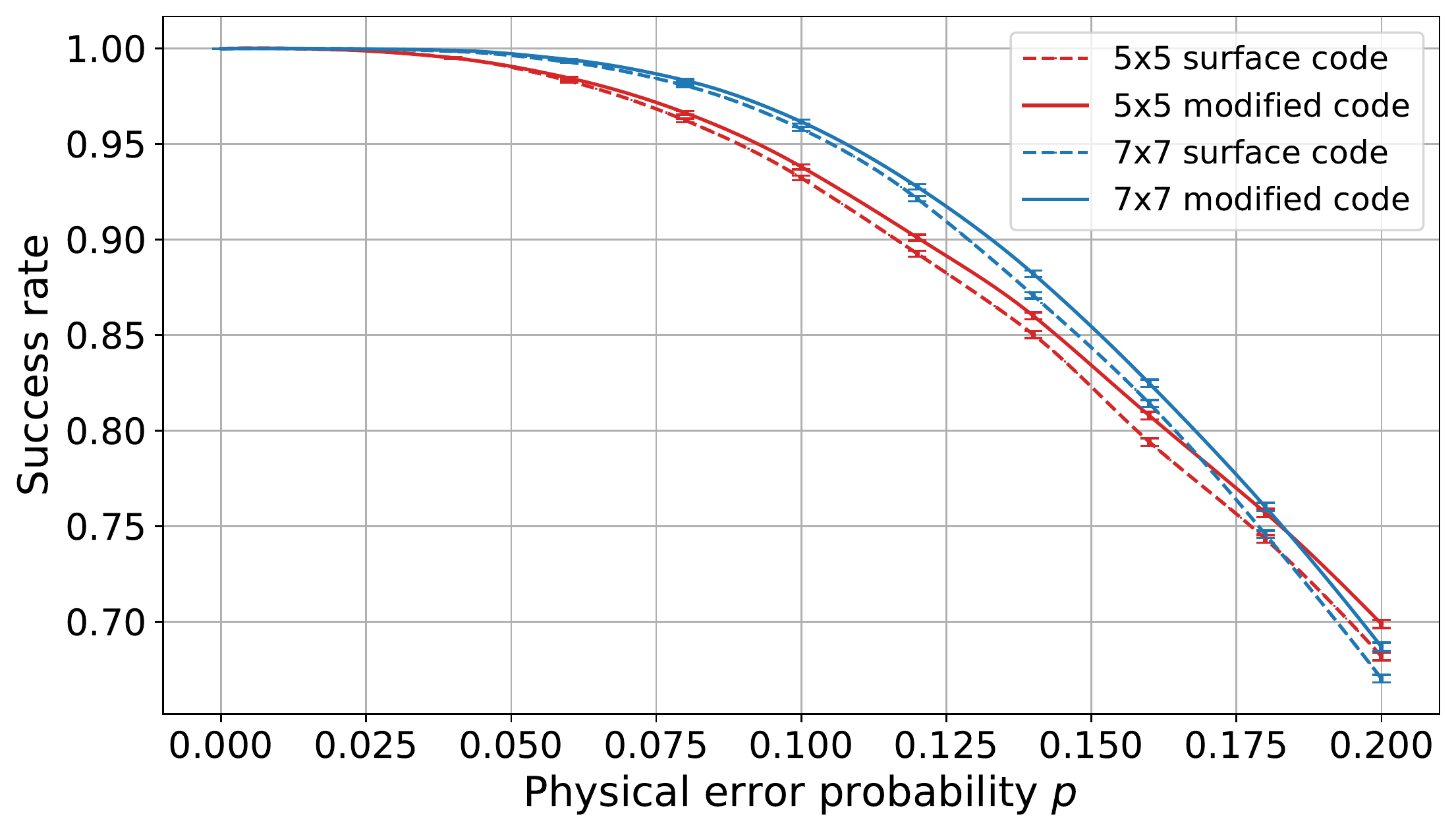}
	\caption{Success rates for the modified surface code and the rotated surface code
	versus the single-qubit error probability $p$.  Here we see an improvement in success probability when using the modified surface code by up to $2\%$.  Interpolating lines are used to distinguish between success rates within a fixed code.  Each point corresponds to 10,000 Monte Carlo samples, and error bars correspond to standard errors.  Choosing relatively small codes allowed us to do the tensor contraction exactly, which ensures that the improvement in performance is not an anomaly due to the choice of bond dimension.
    Our calculations were done using Julia, and code will be made available in the future as the package TensorNetworkCodes.jl, which uses the ITensors.jl package for Julia for tensor network contractions \cite{itensor}.}
\label{fig:code_comparison}
\end{figure}

We will do this by using the tensor-network representation of the rotated surface code and then changing some of the tensors.  The simplest example is to replace the central tensor encoding the logical qubit by the $[[5,1,3]]$ non CSS five-qubit code tensor.  We compare the operator weights for this modified code with that of the rotated surface code.  For example, for a $7\times 7$ code, we see that both codes have the same size $n=49$ and the same distance $d=7$, but the rotated surface code has a degeneracy of the lowest-weight logical operators of $896$, whereas for the modified code, this is only $650$.  This reduction in the number of low-weight logical operators will improve the error correction properties.  Indeed, in Figure \ref{fig:code_comparison}, we see that the modified code does perform somewhat better under depolarizing noise using the tensor-network decoder.  The price we pay is some (four) of the stabilizers are modified to have higher weight, as shown in Figure \ref{fig:new_stabilizers} in appendix \ref{sec:colour}.  An interesting question is if this improvement persists in a fault-tolerant setting, where higher weight stabilizers increase the possibility for additional errors to creep in.  Regardless, this general idea also offers the possibility of optimization strategies to find good codes, e.g., using reinforcement learning.

\section{Discussion}
Our main result is a method that creates general tensor-network codes on graphs.  This also comes with a tensor-network based method to calculate code distances, as well as the full distribution of operator (logical and stabilizer) weights.  For example, we can search for good tensor-network codes by iterating over different contractions of (perhaps randomly chosen) code tensors to find the highest distance code, and then among codes with high distance, we can choose one that has the lowest number of minimal-weight logical operators.  This is precisely what we did to find the modified surface code in section \ref{sec:top}.  By replacing the central code tensor with a five-qubit code tensor with its legs randomly permuted, we found a modification of the rotated surface code with the same distance but fewer minimal-weight logical operators.  We found that this modification led to an improvement in successful error correction probability.  This came at the price of replacing four of the stabilizers by higher-weight stabilizers.  This idea of substituting tensors provides an interesting way to introduce a controlled degree of inhomogeneity into topological codes.  This would allow us to investigate the role of non-locality and adapt codes to physical architectures that support some degree of non-locality.

There are many other open possibilities using tensor-network codes.  One possibility is to start with tensor-network geometries that are efficiently contractible, as these can be decoded efficiently.  Then we can search for error correcting codes that have high thresholds against noise.  This could use a reinforcement learning strategy for example, as this has been successfully applied to designing the layout of surface codes to perform well \cite{NDD19}.  Other questions involve the connection between tensor-network codes and LDPC codes.  For example, it is known that the graph separator of connectivity graphs of quantum codes determines their properties (such as distance, rate and fault-tolerant gate set) \cite{BK21}.  It is interesting to ask whether something similar applies to tensor-network codes when considering the properties of the tensor network graph.

\section*{Note added}
A few days prior to the completion of this paper, a related work appeared on the arXiv \cite{CL21}, where the authors also look at creating larger error correcting codes from smaller code tensors.  This work complements ours nicely, as it includes non-stabilizer codes and a treatment of transversal gates, whereas we focus on stabilizer codes and include a method for distance calculation with tensor networks.

\acknowledgments
The authors are grateful to Ben Brown, Chris Dawson and Rob Harris for useful discussions.
This work was supported by the Australian Research Council Centres of Excellence for Engineered Quantum Systems (EQUS, CE170100009) and the Asian Office of Aerospace Research and Development (AOARD) grant FA2386-18-14027.

\bibliographystyle{unsrt}

\appendix
\section{Code-tensor contraction}
\label{sec:contraction}

To join code tensors to get new stabilizer codes we start by noting that a tensor product of two codes is represented by a product of code tensors:
\begin{equation}
 \begin{split}
	 & T_{12}(L_1\otimes L_2)_{g_1,...,g_{n_1},h_1,...,h_{n_2}} = \\
	 & T_1(L_1)_{g_1,...,g_{n_1}}T_2(L_2)_{h_1,...,h_{n_2}}.
\end{split}
 \end{equation}
 And to understand when we can safely contract indices, we ask when we can contract indices for a general code tensor.  Given code tensors $T(L)_{j_1,...,j_{n}}$ with $n$ physical qubits and $k$ logical qubits.  We can contract two indices to find new tensors describing a new stabilizer code (without loss of generality, we will contract the first two indices).
 \begin{equation}\label{eq:xyz}
	 T_{\mathrm{new}}(L)_{j_3,...,j_{n}} = \sum_{i\in\{0,1,2,3\}}\!\!T(L)_{i,i,j_3,...,j_{n}}.
 \end{equation}
 The only requirement we make now is that the logical degrees of freedom are preserved.  That way, we go from a $[[n,k]]$ code to a $[[n-2,k]]$ code.  In a physical sense, we need that the type II fusion measurement does not measure any logical operators, i.e., we require that we can find representatives of each logical operator that commute with $M_1=\sigma^{1}\otimes\sigma^{1}\otimes\openone...$ and $M_3=\sigma^{3}\otimes\sigma^{3}\otimes\openone...$.
 
 To see how the code is modified by this measurement, we look at which stabilizers $S$ and logicals $L$ survive the measurement.  If we write a stabilizer $S=\sigma^{i_1}\otimes\sigma^{i_2}\otimes\sigma^{i_3}\otimes...$, then if $i_1= i_2$, after the measurement we get that $\pm\sigma^{i_3}\otimes...$ is a new stabilizer on $n-2$ qubits.  The sign depends on the result of the measurement of $M_1$ and $M_3$.  (For simplicity, from here on we will assume the measurement results are both $+1$, but if necessary it is simple to keep track of these minus signs, as we only need them for a set of stabilizer generators.)
 
  \begin{figure*}[ht!]
\includegraphics[width=\textwidth]{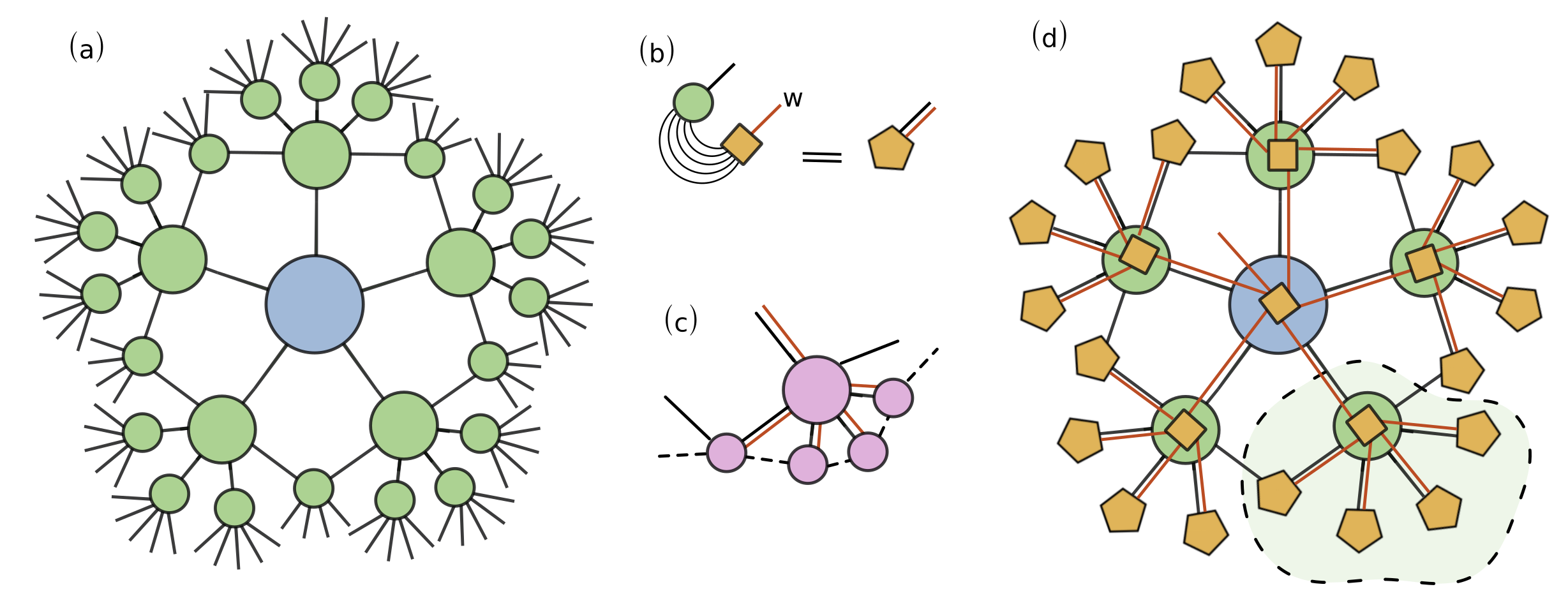}
	\caption{(a) shows an example of a holographic code composed of non-CSS five-qubit code tensors (this holographic code was introduced in \cite{FYH15}).  The central blue tensor corresponds to the five-qubit code tensor, and the remaining green tensors correspond to the purified five-qubit code.  Both codes are described in Table \ref{table:five-qubit}.  The uncontracted legs correspond to physical qubits, so this is a radius-three holographic code.  (b) To calculate the distribution of operator weights, we contract the boundary tensors (green) with the weight tensors $W$ (yellow box) defined in equation (\ref{eq:weighttens}).  Black lines correspond to tensor-network code indices, which have dimension $4$, whereas red lines correspond to weight-tensor indices, which essentially count weights of Pauli operators.  For this example there are five qubits, so $w \in \{0,...,5\}$.  The resultant tensor after contracting is shown as a yellow pentagon.  After performing these contractions at the boundary, we get the tensor network in (d).  To contract this tensor network, we start from the outside and contract inwards.  (c) shows the generic form of the tensor contractions, as seen, e.g., in the shaded region in (d).  (The legs corresponding to dashed lines only arise for higher radius codes.)  All contractions, except the final contraction with the central tensor are of this form (also for higher radius codes).  The crucial point is that the dimensions of the tensor legs are all at most linear in $n$.  This is obvious for the legs from the weight tensors since they count operator weights, and there are only $n$ physical qubits.  For the bond legs (dashed lines in (c)), this was proven in \cite{FHM20}.  Then it follows that the cost of one such contraction is polynomial in $n$.  Furthermore, the number of tensors in the network is also polynomial in $n$, so the number of contractions of the type shown in (c) is also polynomial.  Hence the total complexity is polynomial in $n$.}
\label{fig:holo}
\end{figure*}
 
 \textbf{Case I}:\ neither $M_1$ nor $M_3$ are in the stabilizer group.
 
 In this case for the logical qubits to be preserved after the measurement, it is necessary and sufficient if there are two stabilizers in $\mathcal{S}$, say $S_1=\sigma^{i_1}\otimes\sigma^{i_2}\otimes\sigma^{i_3}\otimes...$ and $S_2=\sigma^{j_1}\otimes\sigma^{j_2}\otimes\sigma^{j_3}\otimes...$ that satisfy
 \begin{equation}
  \begin{split}
   i_1 & \neq i_2\\
   j_1 & \neq j_2
  \end{split}
 \end{equation}
 and, if we take their product, $S_1S_2=\sigma^{k_1}\otimes\sigma^{k_2}\otimes\sigma^{k_3}\otimes...$, then
 \begin{equation}
  k_1\neq k_2.
 \end{equation} 
 This is sufficient because, possibly by taking products with $S_1$ and $S_2$, any other stabilizer and logical operator can be made to have the form $\sigma^{a_1}\otimes\sigma^{a_2}\otimes\sigma^{a_3}\otimes...$ with $a_1=a_2$.
 
 To see that this is necessary, suppose we cannot find such stabilizers.  Then there is at least one undetectable error of the form $\sigma^{l}\otimes\sigma^{l}\otimes\openone...$, which cannot be a stabilizer and hence must be a logical operator.  If this is a logical operator, say logical $X$, then it has a canonically conjugate logical operator, say logical $Z$, which anticommutes with logical $X$.  So this logical $Z$ anticommutes with $\sigma^{l}\otimes\sigma^{l}\otimes\openone...$.  But there is no stabilizer $S$ that we can multiply by $Z$ to make it commute with $\sigma^{l}\otimes\sigma^{l}\otimes\openone...$ because that would require that $S$ anticommutes with $\sigma^{l}\otimes\sigma^{l}\otimes\openone...$, which is impossible because this is a logical operator.  Therefore, logical $Z$ does not survive the fusion measurement.
 
 Next we show that the existence of such $S_1$ and $S_2$ is sufficient for the logical degrees of freedom to be preserved by the measurements.  First, each $S_i$ must anticommute with one or both of $M_1$ and $M_3$.  This follows because $i_1\neq i_2$ and $j_1\neq j_2$.  Furthermore, both $S_1$ and $S_2$ cannot anticommute with both $M_1$ and $M_3$ because that would imply that $S_2S_2$ commutes with both $M_1$ and $M_3$, which is impossible because $k_1\neq k_2$.
 
 Given an operator $O$, which could be another stabilizer, logical operator, or pure error, it may anticommute with one or both of $M_1$ and $M_3$.  But we can choose some product of $S_1$ and $S_2$ that has the same commutation relations with $M_1$ and $M_3$.  Then multiplying $O$ by this product gives a new operator that commutes with both $M_1$ and $M_3$.
 
 Therefore, the remaining $n-k-2$ stabilizer generators (we can choose $S_1$ and $S_2$ to be two of the stabilizer generators) can be chosen to commute with $M_1$ and $M_3$.  Hence, $n-k-2$ stabilizer generators survive the measurement.  The end result is then a stabilizer code on $n-2$ qubits with $k$ logical qubits.
 
 Finally, after contracting the two indices, the values of $T_{\mathrm{new}}(L)_{j_3,...,j_{n}}$ in equation (\ref{eq:xyz}) can only be zero or one, which means we have a valid code tensor.  This is because, for fixed $j_3,...,j_n$, there is at most one operator in each of the logical cosets of the original code of the form $\sigma^i\otimes\sigma^i\otimes\sigma^{j_3}\otimes...$.  If there were two, then their product would be a stabilizer of the form $\sigma^k\otimes\sigma^k\otimes\openone\otimes...$ for some $k\neq 0$.  But this is impossible because this would not commute with $S_1$ and $S_2$.
 
 \textbf{Case II}:\ exactly one of $M_1$ and $M_3$ is in the stabilizer group.
 
Suppose $M_3$ is in the stabilizer group.  Then a simpler argument than the previous case works.  We need only find a stabilizer, say $S_1$, that anticommutes with $M_1$.  More specifically, we need to find a stabilizer $S_1=\sigma^{i_1}\otimes\sigma^{i_2}\otimes\sigma^{i_3}\otimes...$, with $i_1\neq i_2$ and $\{i_1,i_2\}\neq \{0,1\}$ and $\{i_1,i_2\}\neq \{2,3\}$.  These conditions guarantee that $S_1$ will anticommute with $M_1$.  Then, if any of the remaining stabilizers or logical operators anticommute with $M_1$, we can simply multiply by $S_1$ to get an operator that does commute with $M_1$.  It already commutes with $M_3$ because that is a stabilizer.

\begin{figure}[ht!]
\includegraphics[width=\columnwidth]{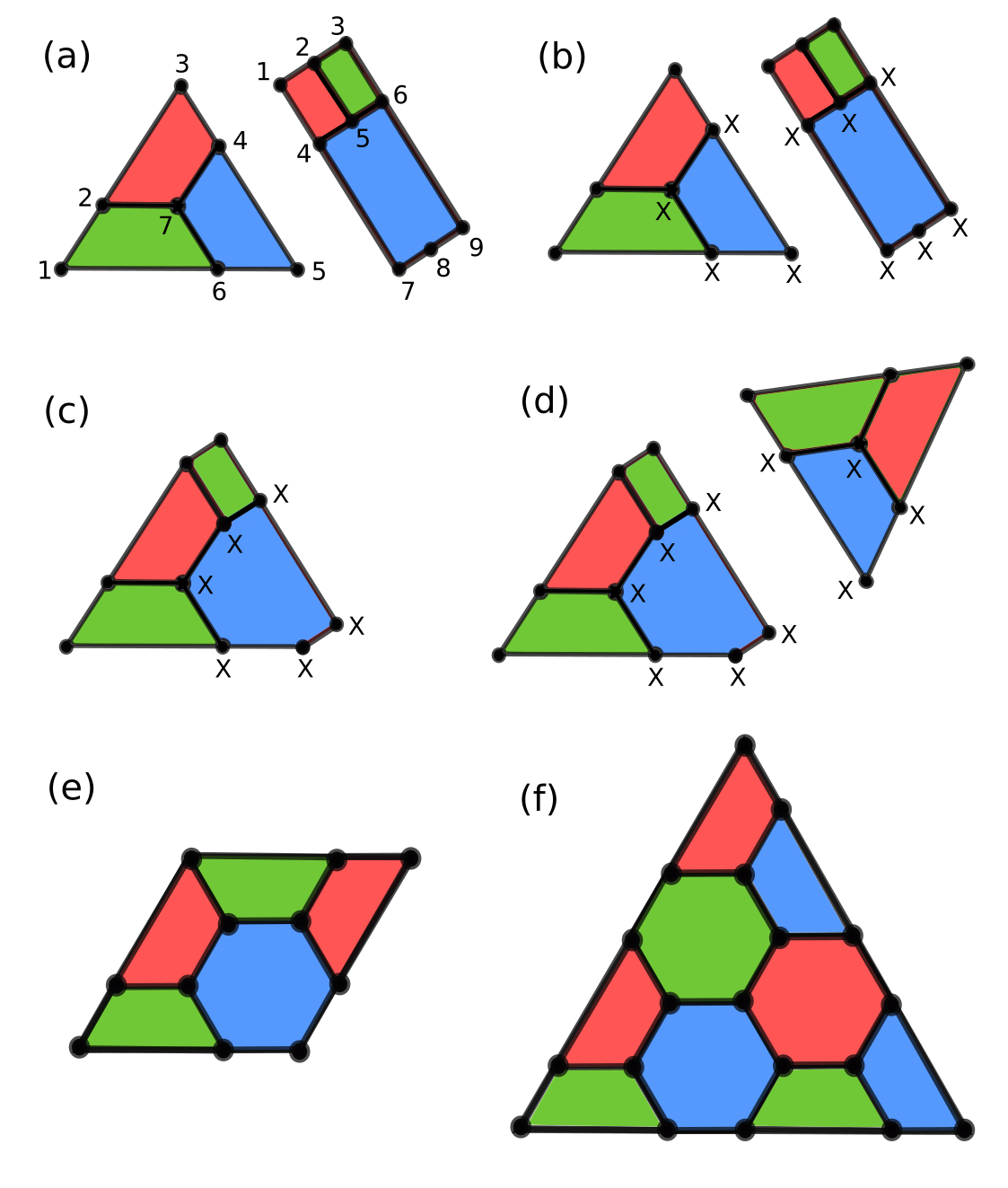}
	\caption{Building up the colour code using the two smaller codes shown in (a).  Stabilizers and logical operators for these codes are listed in Table \ref{table:colour}.  In this case, the pairs of qubit indices contracted are $[3,1]$, $[4,4]$ and $[5,7]$, where the first qubit in each pair belongs to the Steane code (left in (a)) and the second belongs to the nine-qubit code (right in (a)).  (b) shows an example of two stabilizers from each code that match on the pairs of qubits.  (c) shows the resulting stabilizer after contraction.  (d) shows the addition of another Steane code, with two stabilizers shown that match on the qubits that will get contracted.  Any additional Steane codes we contract must have logical $Z$ as an additional stabilizer to reproduce the triangular colour code.  (e) shows the resulting colour code fragment after the contraction in (d).  This procedure can be repeated to build up bigger colour codes, such as the $19$-qubit code shown in (f).}
\label{fig:colour_code}
\end{figure}

Note that we do need to renormalize the tensor
\begin{equation}
	 T_{\mathrm{new}}(L)_{j_3,...,j_{n}} = \frac{1}{2}\sum_{i\in\{0,1,2,3\}}\!\!T(L)_{i,i,j_3,...,j_{n}}.
 \end{equation}

\textbf{Case III}:\ both $M_1$ and $M_3$ are in the stabilizer group.

The logical operators and stabilizers are already in the right form, so there is nothing to do.  Again we need to renormalize
\begin{equation}
	 T_{\mathrm{new}}(L)_{j_3,...,j_{n}} = \frac{1}{4}\sum_{i\in\{0,1,2,3\}}\!\!T(L)_{i,i,j_3,...,j_{n}}.
 \end{equation}

 \begin{table}
\begin{center}
  \begin{tabular}{| c | c c c c c c c | }
    \hline
    (a) & 1 & 2 & 3 & 4 & 5 & 6 & 7 \\ \hline
    $S_1$ & $Z$ & $Z$ & $\openone$ & $\openone$ & $\openone$ & $Z$ & $Z$  \\ \hline
    $S_2$ & $X$ & $X$ & $\openone$ & $\openone$ & $\openone$ & $X$ & $X$  \\ \hline
    $S_3$ & $\openone$ & $Z$ & $Z$ & $Z$ & $\openone$ & $\openone$ & $Z$  \\ \hline
    $S_4$ & $\openone$ & $X$ & $X$ & $X$ & $\openone$ & $\openone$ & $X$  \\ \hline
    $S_5$ & $\openone$ & $\openone$ & $\openone$ & $Z$ & $Z$ & $Z$ & $Z$  \\ \hline
    $S_6$ & $\openone$ & $\openone$ & $\openone$ & $X$ & $X$ & $X$ & $X$  \\ \hline
    $X_1$ & $X$ & $X$ & $X$ & $\openone$ & $\openone$ & $\openone$ & $\openone$  \\ \hline
    $Z_1$ & $Z$ & $Z$ & $Z$ & $\openone$ & $\openone$ & $\openone$ & $\openone$  \\
    \hline
  \end{tabular}
  \quad
  \begin{tabular}{| c | c c c c c c c c c | }
    \hline
    (b) & 1 & 2 & 3 & 4 & 5 & 6 & 7 & 8 & 9 \\ \hline
    $S_1$ & $Z$ & $Z$ & $\openone$ & $Z$ & $Z$ & $\openone$ & $\openone$ & $\openone$ & $\openone$ \\ \hline
    $S_2$ & $X$ & $X$ & $\openone$ & $X$ & $X$ & $\openone$ & $\openone$ & $\openone$ & $\openone$ \\ \hline
    $S_3$ & $\openone$ & $Z$ & $Z$ & $\openone$  & $Z$ & $Z$ & $\openone$ & $\openone$ & $\openone$ \\ \hline
    $S_4$ & $\openone$ & $X$ & $X$ & $\openone$  & $X$ & $X$ & $\openone$ & $\openone$ & $\openone$ \\ \hline
    $S_5$ & $\openone$ & $\openone$ & $\openone$ & $Z$ & $Z$ & $Z$ & $Z$ & $Z$ & $Z$ \\ \hline
    $S_6$ & $\openone$ & $\openone$ & $\openone$ & $X$ & $X$ & $X$ & $X$ & $X$ & $X$ \\ \hline
    $S_7$ & $\openone$ & $\openone$ & $\openone$ & $\openone$ & $\openone$ & $\openone$ & $Z$ & $Z$ & $\openone$ \\ \hline
    $S_8$ & $\openone$ & $\openone$ & $\openone$ & $\openone$ & $\openone$ & $\openone$ & $X$ & $X$ & $\openone$ \\ \hline
    $S_9$ & $\openone$ & $\openone$ & $X$ & $\openone$ & $\openone$ & $X$ & $\openone$ & $\openone$ & $X$ \\ 
    \hline
  \end{tabular}
\end{center}
	\caption{(a) Logical operators and stabilizer generators for the $[[7,1,3]]$ Steane code.  (b) Stabilizer generators for a nine-qubit stabilizer state.  Both codes are illustrated in Figure \ref{fig:colour_code} (a), where qubit numbers in the tables above correspond to numbers in the figure.}
\label{table:colour}
\end{table}
 
\section{Distances for holographic codes}
\label{sec:holo}
Calculating distances for holographic codes using tensor networks is efficient for a similar reason to why tensor-network decoding of holographic codes is exact and efficient \cite{FHM20}.  This is because the corresponding tensor network can be contracted efficiently exactly, which follows roughly because the bond dimension is at most $O(n)$.  The contraction scheme can be understood from Figure \ref{fig:holo}, where the tensor network is contracted from the outside in.  The generic form of each contraction as we go inwards is shown in Figure \ref{fig:holo} (c).  Because there is a fixed number of tensors in each of these contractions, and the legs between them have dimension at most linear in $n$, it follows that the cost of such a contraction is polynomial in $n$.  Then because there are polynomially many such contractions, the total cost is polynomial in $n$.  Therefore, the exact contraction is efficient, so we can efficiently generate the weights of logical operators and stabilizers, which also provides us with the code distance.

\section{More on topological codes}
\label{sec:colour}
\begin{figure*}[ht!]
\includegraphics[width=\textwidth]{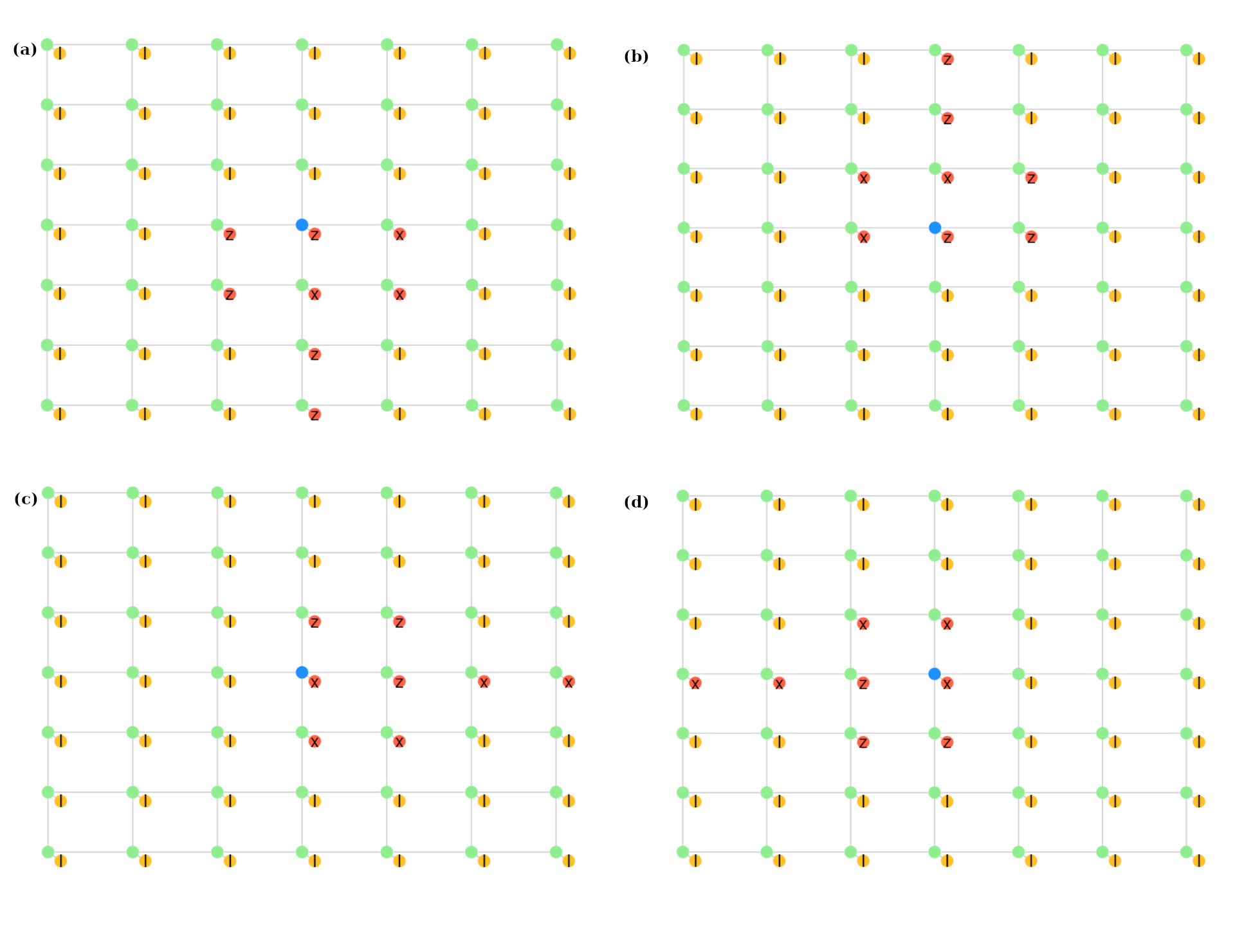}
	\caption{(a), (b), (c) \& (d) show the four new stabilizers in the modified surface code.  All other stabilizers (in plaquettes other than the central four and the boundary) are unchanged.  These new stabilizers for the $7\times 7$ code have weight eight.  For the $5\times 5$ modified code, they have weight $7$.  The weight of these new stabilizers for codes with odd length of side appears to grow as $(L-3)/2 + 6$, so their weight is extensive.  It is worth bearing in mind that this may decrease the effectiveness of the code in a fault-tolerant setting.  Whether this or other similar codes still provide an improvement in performance will depend on the architecture, but a useful goal for tensor-network codes would be to modify codes (or engineer new ones) with bounded-weight stabilizer generators.}
\label{fig:new_stabilizers}
\end{figure*}

The modified surface code described in section \ref{sec:top} has all the local stabilizers of the surface code, except that the central four plaquette stabilizers are replaced by the four displayed in Figure \ref{fig:new_stabilizers}.

To construct the triangular colour code as a tensor-network code, we use two code tensors:\ the Steane code and a nine-qubit code with stabilizers and logical operators listed in Table \ref{table:colour}.  While we find that this is a simple tensor-network code, to understand how the construction works, it is easier to use the graphical description shown in Figure \ref{fig:colour_code}.  Contracting code tensor indices is equivalent to a Type II fusion measurement on those qubits.  After contraction, the stabilizers and logical operators that survive are those that agreed on contracted indices, i.e., on the measured qubits.  In Figure \ref{fig:colour_code} (b), we see examples of two stabilizers that match on the qubits corresponding to indices that will be contracted.  The result is the new stabilizer shown in Figure \ref{fig:colour_code} (c).  Repeating the process with Steane code tensors and the nine-qubit code tensor, we get larger colour codes, such as the $[[19,1,5]]$ colour code in Figure \ref{fig:colour_code} (f).

\end{document}